\documentclass[aps,prd,showpacs,amsmath,amssymb,preprintnumbers]{revtex4}

\newcommand{\vev}[1]{\langle#1\rangle}
\newcommand{\vect}{\left ( \begin{array}{c}}
\newcommand{\evect}{\end{array} \right )}

\usepackage{graphicx}
\usepackage{dcolumn}
\usepackage{bm}
\def\fsl#1{\setbox0=\hbox{$#1$}                 
   \dimen0=\wd0                                 
   \setbox1=\hbox{/} \dimen1=\wd1               
   \ifdim\dimen0>\dimen1                        
      \rlap{\hbox to \dimen0{\hfil/\hfil}}      
      #1                                        
   \else                                        
      \rlap{\hbox to \dimen1{\hfil$#1$\hfil}}   
      /                                         
   \fi}                                         %

\begin{document}

\preprint{\vtop{\hbox{RU06-05-B}
\vskip28pt}}

\title{Inhomogeneity driven by Higgs instability  in gapless superconductor}

\author{Ioannis Giannakis}
\email[E-mail:~] {giannak@summit.rockefeller.edu}
\affiliation{Physics Department, The Rockefeller University,
1230 York Avenue, New York, NY 10021-6399}
\author{Defu Hou}
\email[E-mail:~] {hou@iopp.ccnu.edu.cn}
\affiliation{Institute of Particle Physics, Huazhong Normal University,
Wuhan 430079, China}
\author{Mei Huang}
\email[E-mail:~] {huangm@mail.ihep.ac.cn}
\affiliation{Institute of High Energy Physics, Chinese Academy of Sciences, Beijing 100039, China}
\author{Hai-cang Ren}
\email[E-mail:~] {ren@summit.rockefeller.edu}
\affiliation{Physics Department, The Rockefeller University,
1230 York Avenue, New York, NY 10021-6399}
\affiliation{Institute of Particle Physics, Huazhong Normal University,
Wuhan 430079, China}

\begin{abstract}
The fluctuations of the Higgs and pseudo Nambu-Goldstone fields in the 2SC 
phase with mismatched pairing are described in the nonlinear realization 
framework of the gauged Nambu--Jona-Lasinio model. In the gapless 2SC 
phase,  not only Nambu-Goldstone currents can be spontaneously generated, 
but the Higgs field also exhibits instablity.  The Nambu-Goldstone currents 
generation indicates the formation of the single plane wave LOFF state and 
breaks rotation symmetry, while the Higgs instability favors spatial inhomogeneity 
and breaks translation invariance.  In this paper, we focus on the Higgs instability 
which has not drawn much attention yet. The Higgs instability cannot be removed 
without a long range force, thus it persists in the gapless superfluidity and induces 
phase separation.  In the case of  g2SC state,  the Higgs instability can only be partially 
removed by the electric Coulomb energy. However, it is not excluded that the Higgs
instability might be completely removed  in the charge neutral gCFL phase by the color 
Coulomb energy.

\end{abstract}



\maketitle

\section{Introduction}

Color superconductivity (CSC) \cite{CSC-1,CSC-2} of QCD at high baryon density has been 
an active area of research since 1998. It involves both relativistic field theory and 
statistical mechanics, and many novel properties of this unusual phase of QCD have been 
revealed. For reviews on recent progress see, for example, Ref.~\cite{reviews}. 
The color superconducting phase is expected to reside in the central 
region of compact stars, where the charge neutrality condition as well as $\beta$ 
equilibrium are essential \cite{neutral-QM}. 

The research of charge neutral cold dense quark matter has driven the color 
superconductivity theory into a new territory beyond standard BCS theory.  The requirement 
of charge neutrality condition induces a substantial mismatch between the Fermi surfaces 
of the pairing quarks, which reduces the available phase space for Cooper pairing. 
The results based on conventional BCS framework brings us puzzles. 
For example, in the standard BCS framework, it was found that at moderate mismatch, 
homogeneous gapless superconducting phases \cite{g2SC,gCFL} can be formed. 
However,  gapless superconducting phases exhibit anti-Meissner screening 
effect, i.e., chromomagnetic instability \cite{chromo-ins-g2SC,chromo-ins-gCFL}, 
which is in contradict with the Meissner effect in standard BCS superconductivity. 
 
Superfluidity or superconductivity with mismatched Fermi momenta also appears in 
other systems, e.g., electronic superconductor in a strong external magnetic field
\cite{Sarma, LOFF-orig}, asymmetric nuclear matter \cite{Nuc-M}, and in imbalanced 
cold atomic systems \cite{BP,Wu-Yip}. 
It has been an unsolved old problem how a BCS superconductivity is destroyed 
as the mismatch is increased and what exotic phases may occur before entering the 
normal phase.

It was proposed in the 1960s that, applying a strong external magnetic field in 
an electronic superconductor, the competition between the pair breaking and pair 
condensation would induce an unconventional superconducting phase, i.e., 
Larkin-Ovchinnikov-Fulde-Ferrel (LOFF) state \cite{LOFF-orig}.  
In the LOFF state, the Cooper pair carries a net momentum ${\vec q}$, which
is different from the zero net momentum BCS Cooper pair.  The order parameter of 
the FF state is characterized by a single plane wave: $\Delta ~{\rm e}^{i {\vec q} \cdot {\vec x}}$, 
and the LO state is a superposition of two plane waves or "striped phase": 
$\Delta~ {\rm cos} {\vec q} \cdot {\vec x}$, where $\Delta$ is the amplitude of the gap. 
In general, the structure of the order parameter can be more complicated. 
The LOFF state has still not yet been confirmed experimentally in electronic 
superconduting systems \cite{LOFF-Exp-1, Yang-review} due to the difficulties from 
orbital effects and impurities, and it still remains being persued after 
more than 40 years.

Imbalanced cold atom systems offer another intriguing experimental possibility to understand 
how Cooper pairing is destroyed. By making the populations of atoms in two different hyperfine 
states unequal, one controls the mismatch between their Fermi surfaces. Due to the absence of 
both the orbital effects and impurities, it seems very promising to search for the LOFF state in 
imbalanced cold atom systems. It  has stirred a lot of theoretical interests of discussing
the formation of the LOFF state in mismatched cold atom systems \cite{cold-atom-LOFF}. 
However, recent experiments \cite{LOFF-Exp-atom} in imbalanced cold atom 
systems did not show evidence of the formation of the LOFF state,  rather indicated a 
non-uniform state of phase separation state. 

The momentum-carrying Cooper pairing state was extended to color 
superconductor \cite{LOFF-first} and was further discussed in 
Refs. \cite{LOFF-QM}. However, we started to appreciate the importance of this unconventional
superconducting state only after relating it to chromomagnetic instabilities of 
gapless CSC \cite{FF-Ren}.  Recent studies have demonstrated that the single 
plane wave FF state is unavoidably induced by the instability of the local phase 
fluctuation of the superconducting order parameter, i.e., the Nambu-Goldstone 
boson fields 
\cite{NG-current-hong, NG-current-huang, GHM-gluon,NG-Hashimoto, NG-current-gCFL}. 

Obviously, there should be something missing for our understanding of the pairing breaking state
if we failed to observe the (LO)FF state in the mismatch regime where the magnetic or superfluid 
density instability \cite{Wu-Yip, Pao-Wu-Yip} develops.  In Refs. \cite{Kei-Kenji} and \cite{GHHR},  
a new type of instability, {\it Higgs instability}, was discovered. The Higgs instability, being an 
instability with respect to an inhomogeneous fluctuation of the gap magnitude, may favor non-uniform 
mixed state or LO state more than
the single-plane wave FF state or Nambu-Goldstone current state.

The non-uniform mixed phase has been extensively discussed as one possible candidate of the 
ground state of the asymmetric fermion pairing system \cite{mixed-phase} by comparing the 
free energy,  however, there was no much effort of trying to understand the origin of forming 
a mixed phase.  We try the first step toward this direction by exploring the Higgs instability
of a homogeneous 2SC. The Higgs instability is a straightforward extension 
of the Sarma instability but has not received much attention until recently \cite{GHHR,Kei-Kenji}. 
The Higgs instability refers to the instability of a homogeneous CSC 
state against an inhomogeneous fluctuation of the magnitude of the gap parameter, which is not 
protected by globally imposed constraints, its salient features have been discussed in the short
letter \cite{GHHR}. 

In this paper, we shall mainly display the details behind the arguments made in \cite{GHHR}.  
In Sec.~\ref{sec-gNJL},  we describe the nonlinear realization framework of the 
gauged SU(2) Nambu--Jona-Lasinio(NJL) model in $\beta$-equilibrium. In 
Sec.~\ref{sec-fluc}, we investigate the phase fluctuation as well as the amplitude 
fluctuation, and show that with the increase of mismatch,
the Nambu-Goldstone currents are spontaneously generated, and the Higgs field also becomes 
unstable. In Sec. \ref{sec-general}, we discuss the general features of the Higgs instability. 
The Higgs instability will naturally induce spatial inhomogeneity, however, whether phase 
separation can be developed  in the system has to be determine by its cost of Coulomb 
energy, which is discussed in Sec. ~\ref{sec-com}. We give summary in Sec. ~\ref{sec-summary}.

\section{The gauged SU(2) Nambu--Jona-Lasinio model}
\label{sec-gNJL}

\subsection{The Lagrangian}

To analyze the Higgs instability of the g2SC phase,  
we start from the gauged Nambu--Jona-Lasinio (gNJL) model,
the Lagrangian density has the form of
\begin{eqnarray}
\label{lagr}
{\cal L} = {\bar q}\Big ( i\fsl{D}+\hat{\mu}\gamma^0 \Big ) q  +
   G_S  \Big [(\bar qq)^2+
(\bar qi\gamma^5\vec\tau q)^2\Big ]  +  
G_D \Big [\bar q^Ci\gamma^5\tau_2 \epsilon^{\rho} q\Big ]
\Big [\bar qi\gamma^5\tau_2 \epsilon^{\rho} q^C\Big ] ,
\label{lg}
\end{eqnarray}
with $D_\mu \equiv \partial_\mu - ig A_\mu^{a} T^{a}$.  Here $A_\mu^{a}$ 
are gluon fields,  $T^a=\lambda^a/2$ are the generators 
of $SU(3)_{\rm c}$ gauge group with $a=1, \cdots, 8$. 
In the gNJL model, the gauge fields are external fields and do not contribute to 
the dynamics of the system. The property of the color superconducting phase characterized by the 
diquark gap parameter is determined by the nonperturbative gluon fields, which 
has been simply replaced by the four-fermion interaction in the NJL model.
$G_S$ and $G_D$ are the quark-antiquark coupling constant and the diquark 
coupling constant, respectively.  $q^C=C {\bar q}^T$, ${\bar q}^C=q^T C$ are 
charge-conjugate spinors, where $C=i \gamma^2 \gamma^0$ is the charge conjugation 
matrix (the superscript $T$ denotes the transposition operation). 
The quark field $q \equiv q_{i\alpha}$ with $i=u,d$ and $\alpha=r,g,b$ is a flavor 
doublet and a color triplet, as well as a four-component Dirac spinor. 
${\bf \tau}=(\tau^1,\tau^2,\tau^3)$ are Pauli matrices in the flavor 
space, with $\tau^2$ antisymmetric, and $(\epsilon^{\rho})^{\alpha\beta}$ 
is antisymmetric tensor in color space with $\rho=r,g,b$. 
$\hat{\mu}$ is the matrix of chemical potentials in the color and flavor space.
In $\beta$-equilibrium, the matrix of chemical potentials in the color-flavor 
space ${\hat \mu}$ is given in terms of the quark chemical potential $\mu$, the chemical
potential for the electrical charge $\mu_e$ and the color chemical potential 
$\mu_8$,  
\begin{eqnarray}
\mu_{ij}^{\alpha\beta} = (\mu \delta_{ij} - \mu_e Q_{ij})\delta^{\alpha\beta}
 + \frac{2}{\sqrt{3}} \mu_8 \delta_{ij} (T_8)^{\alpha\beta}.
\end{eqnarray}

In this paper, we shall focus on the color superconducting phase, and neglect the
influence from the chiral condensates by assuming 
$< {\bar q} q >=0$ and $<{\bar q}\gamma^5 {\bf \tau} q>=0$. The is the case when the 
baryon density is sufficiently high.
Introducing  the anti-triplet composite diquark fields 
\begin{eqnarray}
\Delta^{\rho}(x)
&=2iG_D(\bar q^Ci\gamma^5\tau_2 \epsilon^{\rho}q),  \  \
\Delta^{*\rho}(x)
&=-2i G_D (\bar qi\gamma^5\tau_2 \epsilon^{\rho} q^C),
\end{eqnarray}
one obtains the bosonized version of the model for the 2-flavor superconducting phase,  
\begin{eqnarray}
\label{lagr2}
{\cal L}_{2SC} & =  & {\bar q}(i\fsl{D}+\hat \mu \gamma^0)q 
-\frac{1}{4G_D} \Delta^{*\rho}\Delta^{\rho}
 \nonumber \\
 & + & \frac{i\Delta^{*\rho}}{2}[\bar q^Ci\gamma^5\tau_2\epsilon^{\rho} q]
-\frac{i\Delta^{\rho}}{2}[\bar q i\gamma^5\tau_2 \epsilon^{\rho}q^C].
 \label{lagr-2sc}
\end{eqnarray}

In the Nambu-Gor'kov space, 
\begin{equation}
  \Psi = \left(\begin{array}{@{}c@{}} q \\ q^C \end{array}\right),
\end{equation}
the Lagrangian (\ref{lagr-2sc}) simplifies to
\begin{equation}
{\cal L}_{2SC}=-\frac{\Delta^{*\rho}\Delta^{\rho}}{4G_D}+
\frac 12\bar\Psi \left[{\cal S}\right]^{-1} \Psi.
\label{lg-NG}
\end{equation}
The inverse of the quark propagator in momentum space is defined as
\begin{equation}
\left[{\cal S}(P)\right]^{-1} = \left(\begin{array}{cc}
\left[G_0^{+}(P)\right]^{-1} & \Delta^- \\
\Delta^+ & \left[G_0^{-}(P)\right]^{-1}
\end{array}\right),
\label{prop}
\end{equation}
with the off-diagonal elements
\begin{equation}
  \Delta^- \equiv -i \tau_2\epsilon^{\rho} \gamma_5 \Delta^{\rho}, \qquad
  \Delta^+ \equiv i \tau_2\epsilon^{\rho} \gamma_5 \Delta^{*\rho},
\end{equation}
and the free quark propagators $G_0^{\pm}(P)$ taking the form of
\begin{equation}
\left[G_0^{\pm}(P)\right]^{-1} =
 \gamma^0 (p_0 \pm \hat{\mu}) - \vec{\gamma} \cdot \vec{p},
 \label{freep}
\end{equation}
with the 4-momenta denoted by capital letters, e.g., $P=(p_0,\vec{p})$. 
In terms of the Pauli matrices with respect to NG indexes,
\begin{equation} 
\rho_1=\left(\begin{array}{cc} 0 & 1 \\ 1 & 0 \end{array}\right), 
\qquad
\rho_2=\left(\begin{array}{cc} 0 & -i \\ i & 0 \end{array}\right), 
\qquad
\rho_3=\left(\begin{array}{cc} 1 & 0 \\ 0 & -1 \end{array}\right), 
\end{equation}
the inverse propagator (\ref{prop}) can be cast into a compact form
\begin{equation}
\left[{\cal S}(P)\right]^{-1}=\gamma^0(p_0+\rho_3\hat\mu)-\vec\gamma\cdot\vec p
-\rho_+\Delta^-+\rho_-\Delta^+
\end{equation}
with $\rho_{\pm}=(\rho_1\pm i\rho_2)/2$.

\subsection{Nonlinear realization framework of the gNJL model}

The superconducting state
is characterized by the order parameter  $\Delta(x)$, which is a complex scalar field 
and has the form of $\Delta (x) = |\Delta(x)| e^{i\varphi (x)}$, with $|\Delta| $ the amplitude 
and $\varphi$ the phase of the gap order parameter. For a homogeneous condensate, 
$\Delta(x)$ is a spatial constant. The fluctuations of the phase give rise to 
the pseudo Nambu-Goldstone boson(s), while that of the amplitude to the Higgs field, following 
the terminology of the electroweak theory. 
Stimulated by the role of the phase fluctuation in the unconventional superconducting
phase \cite{HTSC} in condensed matter, we follow Ref. \cite{NonLinear} to formulate the 2SC phase 
in the nonlinear realization framework in order to take into account naturally the contribution 
from the phase fluctuation or pseudo Nambu-Goldstone current(s). 

In the 2SC phase, the color symmetry $G=SU(3)_c$  breaks to $H=SU(2)_c$.
The generators of the residual $SU(2)_c$ symmetry H are 
$\{S^a=T^a\}$ with $a=1,2,3$ and the broken generators $\{X^b=T^{b+3}\}$ 
with $b=1, \cdots, 5$. (More precisely, the last broken 
generator is a combination of $T_8$ and the generator ${\bf 1}$ 
of the global $U(1)$ symmetry of baryon number conservation, 
$B \equiv ({\bf 1} + \sqrt{3} T_8)/3$ of generators of
the global $U(1)_B$ and local $SU(3)_c$ symmetry.
)

The coset space $G/H$ is parameterized by the group elements
\begin{equation} \label{phase}
{\cal V}(x) \equiv \exp \left[ i \left( \sum_{a=4}^8 \varphi_a(x) T_a \right) \right]  \,\,,
\end{equation}
here operator ${\cal V}$ is unitary, and ${\cal V}^{-1} = {\cal V}^\dagger$
and $\varphi_a (a=4,\cdots,7)$ and $\varphi_8$ are five Nambu-Goldstone 
diquarks, and we have not consider the topologically nontrivial case and 
therefore ${\cal V}(x)$ can be expanded uniformly according to the powers 
of $\varphi$'s. In fact, ${\cal V}(x)$ is alway topologically trivial 
for a configuration of $\varphi$'s that has a finite energy because of 
the trivial homotopy group $\pi_2(SU(3)/SU(2))$ \cite{atiyah}.

Introducing a new quark field $\chi$, which is connected with the original 
quark field $q$ in Eq. (\ref{lagr-2sc}) through a nonlinear transformation,
\begin{equation} \label{chi}
q = {\cal V}\, \chi
\,\,\,\, , \,\,\,\,\,
\bar{q} = \bar{\chi}\, {\cal V}^\dagger\,\, ,
\end{equation}
and the charge-conjugate fields transform as
\begin{equation}
q_{C} = {\cal V}^* \, \chi_{C}
\,\,\,\, , \,\,\,\,\,
\bar{q}_{C} = \bar{\chi}_{C} \, {\cal V}^T\,\, .
\end{equation}
The advantage of transforming the quark fields is
that this preserves the simple structure of the terms coupling
the quark fields to the diquark sources,
\begin{equation}
\bar{q}_{C}\, \Delta^+ \, q
\equiv \bar{\chi}_{C}\, \Phi^+ \, \chi
\,\,\,\, , \,\,\,\,\,
\bar{q}\, \Delta^- \, q_{C}
\equiv \bar{\chi} \, \Phi^- \, \chi_{C} \,\, .
\end{equation}

In the Nambu-Gor'kov space of the new spinors
\begin{equation}
X \equiv \left( \begin{array}{c} 
                    \chi \\
                    \chi_{C} 
                   \end{array}
            \right) \,\,\, , \,\,\,\,
\bar{X} \equiv ( \bar{\chi} \, , \, \bar{\chi}_{C} ),
\end{equation}
the nonlinear realization of the original Lagrangian density
Eq.~(\ref{lg-NG}) takes the form of
\begin{equation}
{\cal L}^{nl}_{2SC} \equiv - \frac{\Phi^+\Phi^-}{4 G_D} \,\, + \frac{1}{2}
\bar{X} \,  {\cal S}_{nl}^{-1} \, X ,
\label{lagr-nl}
\end{equation}
with
\begin{equation}
{\cal S}_{nl}^{-1} \equiv 
\left( \begin{array}{cc}
            [G^+_{0,nl}]^{-1} & \Phi^- \\
             \Phi^+ & [G^-_{0,nl}]^{-1}
       \end{array} \right)\,\, .
\end{equation}
Here the explicit form of the free propagator for the new quark field is
\begin{eqnarray}
[G^+_{0,nl}]^{-1} & = & i\, \fsl{D} + {\hat \mu} \, \gamma_0 + \gamma_\mu \, V^\mu, 
\end{eqnarray}
and
\begin{eqnarray}
[G^-_{0,nl}]^{-1} & = & i\, \fsl{D}^T - {\hat \mu} \, \gamma_0 + \gamma_{\mu} \, V_C^\mu .
\end{eqnarray}
Comparing with the free propagator in the original Lagrangian density, 
the free propagator in the non-linear realization framework naturally takes
into account the contribution from the Nambu-Goldstone currents or 
phase fluctuations, i.e.,
\begin{eqnarray}
V^\mu & \equiv & {\cal V}^\dagger \, \left( i \, \partial^\mu \right) \, {\cal V}, \nonumber \\
V^\mu_C & \equiv & {\cal V}^T \, \left( i \, \partial^\mu \right) \, {\cal V}^*,
\end{eqnarray}
which is the $N_c N_f \times N_c N_f$-dimensional
Maurer-Cartan one-form introduced in Ref. \cite{NonLinear}.
The linear order of the Nambu-Goldstone currents 
$V^\mu$ and $V_C^\mu $ has the explicit form of
\begin{eqnarray}
V^\mu & \simeq &  - \sum_{a=4}^8 
\left( \partial^\mu  \varphi_a \right)\, T_a  \,\, , \\
V_C^\mu & \simeq &   \sum_{a=4}^8 
 \left(\partial^\mu \varphi_a \right) \, (T_a)^T  \,\, .
\end{eqnarray}

The advantage of the non-linear realization framework Eq. (\ref{lagr-nl}) is
that it can naturally take into account the contribution from the phase fluctuations 
or Nambu-Goldstone currents. 
The task left is to find the correct ground state by exploring the stability 
against the fluctuations of the magnitude and the phases of the order parameters. 
The free energy $\Omega(V_{\mu},\Phi, \mu,\mu_8, \mu_e)$ can be evaluated 
 directly and it  takes the form of
\begin{eqnarray}
\Omega_{nl}(V_{\mu}, \Phi, \mu, \mu_8, \mu_e) = 
- \frac{1}{2} T\sum_n\int\frac{d^3\vec{p}}{(2\pi)^3} 
{\rm Tr} \ln ( [{\cal S}_{nl}(P)]^{-1}) + \frac{\Phi^2}{4 G_D}.
\end{eqnarray}

\section{The fluctuation of the Higgs and Nambu-Goldstone fields}
\label{sec-fluc}

\subsection{Expansion the free-energy in terms of the BCS Cooper-pairing state}

In the 2SC phase, the color symmetry $SU(3)_c$  is spontaneously broken to $SU(2)_c$
and diquark field obtains a nonzero expectation value. 
Without loss of generality, one can always assume that diquark condenses in the
anti-blue direction, i.e., only red and green quarks participate the Cooper pairing, 
while blue quarks remains as free particles. The ground state of the 2SC phase is 
characterized by  $\vev{\Delta^{3}}\equiv \Delta$, and $\vev{\Delta^{1}}=0$,
$\vev{\Delta^{2}}=0$. 

Considering the fluctuation of the order parameter,  the diquark condensate can be 
parameterized as
\begin{equation}
\vect\Delta^1(x)\\ \Delta^2(x)\\ \Delta^3(x)\evect =
\exp \left[ i \left( \sum_{a=4}^8 \varphi_a(x) T_a
 \right) \right] \vect0\\0\\ \Delta + H(x) \evect  
\equiv {\cal V}(x) \Phi^{\rho}(x),
\label{II}
\end{equation}
where $\Phi^{\rho}(x)= (0, \, 0, \, \Delta + H(x))$ is the diquark field in the nonlinear realization framework, 
$\varphi_a, \varphi_8$ are Nambu-Goldstone bosons, and $H$ is the Higgs field.

Expanding the diquark field $\Phi^{\rho}$ around the ground state: $\Phi^{\rho}= (0, \, 0, \, \Delta)$,
the free-energy of the system takes the following expression as
\begin{eqnarray}
\Omega_{nl} = \Omega_M +\Omega_{NG} + \Omega_{H}.
\label{free-energy-NG}
\end{eqnarray}
There are three contributions to the free-energy, the mean-field approximation free-energy
part $\Omega_M$ which has the form of
\begin{equation}
\Omega_M= - \frac{T}{2} \sum_n\int\frac{d^3\vec{p}}{(2\pi)^3} {\rm Tr} \ln ( [{\cal S}_M(P)]^{-1}) + \frac{\Delta^2}{4 G_D},
\end{equation}
the free-energy from the Higgs field $\Omega_H$ has the form of
\begin{equation}
\Omega_H = \frac{T}{2}\sum_{k_0}\int\frac{d^3\vec k}{(2\pi)^3}H^*(K)\Pi_{H}(K)H(K)
\label{higgs}
\end{equation}
with
\begin{equation}  
\Pi_{H}(K)=\frac{1}{2G_D} - \frac{T}{2}\sum_{p_0}\int\frac{d^3\vec{p}}{(2\pi)^3} {\rm Tr}\Big [ 
{\cal S}_M(P+K)\, 
\left(\begin{array}{cc}
0 & i \tau_2\epsilon^{3} \gamma_5 \\
-i \tau_2\epsilon^{3} \gamma_5 & 0
\end{array}\right) \, {\cal S}_M(P) \, 
\left(\begin{array}{cc}
0 & i \tau_2\epsilon^{3} \gamma_5 \\
-i \tau_2\epsilon^{3} \gamma_5 & 0
\end{array}\right)\Big],
\end{equation}
and the free-energy from the Nambu-Goldstone currents $\Omega_{NG}$ has the form of
\begin{eqnarray}
\Omega_{NG} & = &  - \frac{T^2}{4} \sum_{k_0} \sum_{p_0}\int\frac{d^3\vec{k}}{(2\pi)^3}
\frac{d^3\vec{p}}{(2\pi)^3} \nonumber \\
& & {\rm Tr} \Big [ {\cal S}_M(P+K)\, 
\left(\begin{array}{cc}
\omega^\mu(-K)\gamma_\mu & 0 \\
0 & \omega_C^\mu(-K)\gamma_\mu
\end{array}\right) \,
{\cal S}_M(P) \, 
\left(\begin{array}{cc}
\omega^\mu(K)\gamma_\mu & 0 \\
0 & \omega_C^\mu(K)\gamma_\mu
\end{array}\right) \, \Big ],
\label{goldstone}
\end{eqnarray}
with 
\begin{eqnarray}
\omega^\mu(K) & = & g \, A^\mu_a(K) \, T_a - V^{\mu}(K) \,\, , \\
\omega_C^\mu(K) & =& - g \, A^\mu_a(K) \, T_a^T + V_C^{\mu}(K)\,\, .
\end{eqnarray}
where the inverse propagator ${\cal S}_M^{-1}$ takes the form of
\begin{eqnarray}
\left[{\cal S}_M(P)\right]^{-1} & = & \left(\begin{array}{cc}
\left[G_0^{+}(P)\right]^{-1} & i \tau_2\epsilon^{3} \gamma_5 \Delta \\
-i \tau_2\epsilon^{3} \gamma_5 \Delta & \left[G_0^{-}(P)\right]^{-1}
\end{array}\right) \nonumber \\
& = & \gamma^0(p_0+\rho_3\hat\mu)-\vec\gamma\cdot\vec p+\Delta\rho_2\tau_2\epsilon^3\gamma_5.
\label{prop-0}
\end{eqnarray}
The quasi-quark propagator at mean-field approximation has the form of 
\begin{equation} 
{\cal S}_M = \left(\begin{array}{cc}
G^{+} & \Xi^{-} \\
\Xi^{+} & G^{-}
\end{array}\right).
\label{quarkpropagator}
\end{equation}
and its explicit expression of the Nambu-Gorkov components of ${\cal S}_M$
has been derived in Ref. ~\cite{chromo-ins-g2SC}. The weak coupling approximation of 
${\cal S}_M$ can be found in the appendix A of this paper.

The Matsubara self-energy functions $\Pi_H(K)$ of the Higgs field and that of the Goldstone fields
(obtained after the sum over $p_0$ and integral over $\vec p$ in Eq.~(\ref{goldstone}))
can be continuated to real frequency following the standard procedure. Because of the gapless 
excitations, the values of these functions at zero frequency and zero mometum depends the order 
of the limit. In this work, we shall restrict our attention to static fluctuations only which 
amounts to replace $H(K)$, $\vec A(K)$ and $\varphi(K)$ of Eqs.~(\ref{higgs}) and (\ref{goldstone})
by $\sqrt{T}H(\vec k)\delta_{k_0,0}$, $\sqrt{T}\vec A(\vec k)\delta_{k_0,0}$ and 
$\sqrt{T}\varphi(\vec k)\delta_{k_0,0}$. Thus the long wavelength limit of the 
self-energy functions discussed below corresponds to the limit $\lim_{\vec k\to 0}\lim_{k_0\to 0}$.
\subsection{Free-energy at mean-field approximation and Sarma instability}

In the mean-field approximation, the free-energy  for $u, d$ quarks in 
$\beta$-equilibrium  takes the form \cite{g2SC}:
\begin{eqnarray} 
\Omega_M = 
 \frac{\Delta^2}{4G_D}
-\sum_{A} \int\frac{d^3 p}{(2\pi)^3} \left[E_{A}
+2 T\ln\left(1+e^{-E_{A}/T}\right)\right] .
\label{pot-m}
\end{eqnarray} 
Here we did not take into account the contribution of free electrons.
The sum over $A$ runs over all (6 quark and 6 antiquark)
quasi-particles. 

At zero temperature, the mean-field free-energy has the expression of 
\begin{eqnarray}
\Omega_M &=& 
\frac{\Delta^2}{4G_D} 
-\frac{\Lambda^4}{2\pi^2} 
-\frac{\mu_{ub}^4}{12 \pi^2}
-\frac{\mu_{db}^4}{12 \pi^2} \nonumber \\
&-& 2 \int_{0}^{\Lambda} \frac{p^2 d p}{\pi^2} 
\left(\sqrt{(p+ \bar{\mu})^2+\Delta^2}
+\sqrt{(p-\bar{\mu})^2+\Delta^2}\right)\nonumber \\
&-& 2\theta \left(\delta\mu-\Delta\right)
\int_{\mu^{-}}^{\mu^{+}}\frac{p^2 d p}{\pi^2}\Big(
\delta\mu-\sqrt{(p-\bar{\mu})^2+\Delta^2}
\Big).
\label{pot-2sc}
\end{eqnarray}

As we already knew that, with the increase of mismatch, the ground state
will be in the gapless 2SC phase when $\Delta < \delta\mu$, the thermodynamical 
potential of which is given by
\begin{equation}
\Omega_M\simeq \Omega_M^{(0)}+\frac{2\bar\mu^2}{\pi^2}
\Big(\ln\frac{\delta\mu+\sqrt{\delta\mu^2-\Delta^2}}{\Delta_0}
-\delta\mu\sqrt{\delta\mu^2-\Delta^2}+\delta\mu^2\Big)
\label{omegaM}
\end{equation}
where $\Omega_M^{(0)}$ is the normal phase thermodynamic potential. 
$\Delta_0$ the solution to the gap equation in the absence of mismatch, $\delta\mu=0$. The solution to the gap equation reads
\begin{equation}
\Delta=\sqrt{\Delta_0(2\delta\mu-\Delta_0)}.
\label{gapsol}
\end{equation}

The gapless phase is in principle a metastable Sarma state \cite{Sarma}, i.e., the free-energy
is a local maximum with respect to the gap parameter $\Delta$. We have 
\begin{equation}
\Big(\frac{\partial^2\Omega_M}{\partial\Delta^2}\Big)_{\bar\mu,\delta\mu}
=\frac{4\bar\mu^2}{\pi^2}\Big(1-\frac{\delta\mu}{\sqrt{\delta\mu^2-\Delta^2}}\Big).
\label{sarmains}
\end{equation}
The weak coupling approximation is employed in deriving Eqs.~(\ref{omegaM}) and (\ref{sarmains})
from Eq.~(\ref{pot-2sc}), which assumes that $\Delta_0$, $\Delta$ and $\delta\mu$ are much smaller 
than $\mu$ and $\Lambda-\mu$. The same approximation will be applied throughoutthe paper.

\subsection{Nambu-Goldstone currents generation and the LOFF state}

The quadratic action of the Goldstone modes in the long wavelength limit can be written down 
with the aid of the Meissner masses evaluated in Ref.\cite{chromo-ins-g2SC}. We find that
\begin{equation}
\label{NG-free-energy}
\Omega_{NG}=\frac{1}{2}\int d^3\vec r\sum_{a=1}^8 m_a^2 ({\vec {\bf A}}^a 
-\frac{1}{g}\, {\vec \triangledown}  \varphi^a)( {\vec {\bf A}}^a- \frac{1}{g}\, {\vec \triangledown}  \varphi^a) 
+ higher \, orders \, .
\end{equation}
where $m_1=m_2=m_3=0$, 
\begin{equation}
m_4^2=m_5^2=m_6^2=m_7^2=\frac{g^2\bar\mu^2}{3\pi^2}\Big[\frac{\Delta^2-2\delta\mu^2}{2\Delta^2}
+\theta(\delta\mu-\Delta)\frac{\delta\mu\sqrt{\delta\mu^2-\Delta^2}}{\Delta^2}\Big]
\end{equation}
and 
\begin{equation}
m_8^2=\frac{g^2\bar\mu^2}{9\pi^2}\Big[1-\frac{\delta\mu\theta(\delta\mu-\Delta)}{\sqrt{\delta\mu^2-\Delta^2}}\Big].
\end{equation}
It was found that at
zero temperature, with the increase of mismatch,  for five gluons with $a=4,5,6,7,8$ 
corresponding to broken generator of $SU(3)_c$, their Meissner screening mass squares
become negative \cite{chromo-ins-g2SC}. This indicates the development of 
the condensation of 
\begin{equation}
\sum_{a=4}^{8} <{\vec {\bf A}}^a- \frac{1}{g}\, {\vec \triangledown}  \varphi^a> \neq 0.
\end{equation}
It can be interpreted as the spontaneous generation of Nambu-Goldstone currents 
$\sum_{a}^{ 8}  <{\vec \triangledown}  \varphi^a> \neq 0$ \cite{NG-current-huang}, or gluon condensation
$\sum_{a=4}^{8}  <{\vec {\bf A}}^a> \neq 0$ \cite{GHM-gluon}. It can also be interpreted as
a colored-LOFF state with the plane-wave order parameter 
\begin{equation}
\Delta(x)= \Delta {\rm e}^{ i \sum_{a=4}^{8}  {\vec \triangledown}  \varphi^a \cdot {\vec {\bf x}}}.
\end{equation}

\subsection{Higgs instability}
\label{sub-Higgs-ins}

In the previous two subsections, we discussed the two known instabilities
induced by mismatch, 
i.e., the Sarma instability and chromomagnetic instability, respectively. 
In this subsection, we focus on another instability, which is related to the Higgs field, and 
we call this instability "Higgs instability".

The free-energy from the Higgs field can be evaluated and takes the form of
\begin{eqnarray} 
\label{free-energy-H}
\Omega_{H} & =&  \frac{T}{2}\sum_{k_0}\int\frac{d^3\vec{k}}{(2\pi)^3} 
 H^*(\vec k)\Pi_{H}(k) H(\vec k).
\end{eqnarray}

\begin{figure}[ht]
\vspace*{-3truecm}
\includegraphics[width=16cm]{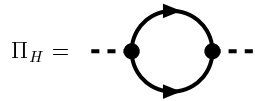}
\vspace*{-16truecm}
\caption{\label{Higgs-fig}The Feynman diagrams for the Higgs polarization function $\Pi_{H}$.}
\end{figure}

Evaluating the one-loop quark-quark bubble $\Pi_H(k)$ shown in Fig. \ref{Higgs-fig}, we obtain that  
\begin{equation}
\Pi_H(k)=\frac{2\bar\mu^2}{\pi}I(k|\delta\mu)+\frac{\bar\mu^2}{2\pi\Delta^2}k^2J(k|\delta\mu),
\label{higgsenergy}
\end{equation}
where the functions $I(k|\delta\mu)$ and $J(k|\delta\mu)$ are given by
\begin{equation}
I(k|\delta\mu) = \Delta^2T \sum_{n}\frac{1}{\sqrt{(\omega_n+i\delta\mu)^2+\Delta^2}}
\int_{-1}^1dx\frac{1}{(\omega_n+i\delta\mu)^2+\Delta^2+\frac{1}{4}k^2x^2},
\label{Iintegral}
\end{equation}
\begin{equation}
J(k|\delta\mu) = \Delta^2T \sum_{n}\frac{1}{\sqrt{(\omega_n+i\delta\mu)^2+\Delta^2}}
\int_{-1}^1dx\frac{x^2}{(\omega_n+i\delta\mu)^2+\Delta^2+\frac{1}{4}k^2x^2}.
\label{Jintegral}
\end{equation}
with ${\rm Re}\sqrt{(\omega_n+i\delta\mu)^2+\Delta^2}>0$.
Here, the summation over $n$ is the frequency summation at finite temperature field theory, with
$\omega_n=(2n+1)\pi T$.  
At T=0, both functions can be simplified as shown in the appendices A and B. Their power series expansion yield
\begin{equation}
\Pi_H(k)=A_H+B_Hk^2
\end{equation}
with
\begin{equation}
A_H = \left(\frac{\partial^2\Omega_M}{\partial\Delta^2}\right)_{\delta\mu}
=\frac{4\bar\mu^2}{\pi^2}\Big(1-\frac{\delta\mu}{\sqrt{(\delta\mu)^2-\Delta^2}}\Big),
\label{AH}
\end{equation}
\begin{equation}
B_H=\frac{2\bar\mu^2}{9\pi^2\Delta^2}\Big[1-\frac{(\delta\mu)^3}{((\delta\mu)^2-\Delta^2)^{\frac{3}{2}}}
\Big].
\label{BH}
\end{equation}
It follows from Eqs. (\ref{AH}) and  (\ref{BH}) that the Higgs field 
becomes unstable in the gapless phase when $\delta\mu > \Delta$, 
$A_H$ in the gapless phase is shown in Fig. \ref{Higgs-f-fig} by the red dash-dotted line.
The Higgs instability was also considered in Ref. \cite{Kei-Kenji}, where 
it was called as "amplitude instability". It has to be pointed out that we got 
different expressions for the coefficients of the gradient term.
For $k>>\Delta$, we have 
\begin{equation}
\Pi_H(k)\simeq \frac{\bar\mu^2}{2\pi^2}\Big(2\ln\frac{k}{\Delta}-2
-\ln\frac{\delta\mu-\sqrt{\delta\mu^2-\Delta^2}}{\delta\mu+\sqrt{\delta\mu^2-\Delta^2}}\Big).
\end{equation}
Therefore the Higgs instability disappears for sufficiently large momentum.
The form factor $\Pi_H(k)$ for arbitrary momentum is plotted using red dashed line
in Fig. \ref{Higgs-k-fig} for a typical value of the mismatch parameter
$\Delta/\delta\mu=1/2$. We notice that the 
Higgs instability becomes stronger for nonzero momentum. 

Near the transition temperature, $\Delta<<T$, we may drop $\Delta$'s inside 
the square roots and in the  denominators on RHS of Eqs.~(\ref{Iintegral}) 
and (\ref{Jintegral}). 
The Higgs self-energy function (\ref{higgsenergy}) goes to the expression derived from 
the Ginzburg-Landau theory \cite{Type-trans}.
The Ginzburg-Landau free energy for g2SC can be easily obtained from that of an electronic 
superconductor by replacing the Fermi momentum $k_F$ with $\bar\mu$, the Fermi velocity
with one and multiplying the whole expression by four (four pairing configurations 
per momentum for g2SC versus one for the electronic SC).

We would like to make two remarks before concluding this section: 1) The combination of 
$\vec A-\frac{1}{g}\vec\nabla\varphi$ of Eq.~(\ref{NG-free-energy}) is by no means generic for a 
nonrelativistic system. The general expression of $\Omega_{NG}$ beyond the long wavelength limit takes the form 
\begin{equation}
\Omega_{NG}=\frac{1}{2}\int \frac{d^3\vec k}{(2\pi)^3}\sum_{a=1}^8 \left[ 
\Pi_a^\perp(k){\vec {\bf A}}_\perp^{a*}(\vec k)\cdot{\vec {\bf A}}_\perp^a(\vec k) 
+\Pi_a^{||}(k)({\vec {\bf A}}_{||}^{a*}+\frac{i}{g}\vec k\varphi^{a*}(\vec k))
\cdot({\vec {\bf A}}_{||}^a(\vec k)-\frac{i}{g}\vec k\varphi^a(\vec k)) 
+ higher \, orders \, \right].
\end{equation}
in momentum representation, where $\vec A_{||}$($\vec A_\perp$) is parallel(perpendicular) 
to $\vec k$. Except that $\Pi_a^\perp(0)=\Pi_a^{||}(0)$, they may be quite different for 
$\vec k\neq 0$, so are the instability domains with respect to $k$.
2) At $T=0$, the dependence of the thermodynamic function on the diquark field is highly nonlinear and the 
gradient term is not Ginzburg-Landua like, i.e., $\vec\nabla\Delta^{\rho*}\cdot\vec\nabla\Delta^\rho$. 
There are no simple relations between the coefficient $B_H$ and the Meissner mass squares. This may 
explain the difference between our $B_H$ and that of \cite{Kei-Kenji}.

\section{A General View of the Higgs Instability}
\label{sec-general}

In this section, we shall provide a general view
of the Higgs instability in relation to the Sarma instability as well as
the Coulomb energy in relation to the charge neutrality constraint.
Consider a system described by the Hamiltonian ${\cal H}$ with a spontaneous
symmetry breaking. The thermodynamic potential is given by
\begin{equation}
\Omega = -T\ln{\rm Tr}\exp\left(-\frac{{\cal H}-\sum_{j=1}^m\mu_jQ_j}
{T}\right)
\end{equation}
where $Q$'s are $m$ conserved charges with $\mu$'s the corresponding
chemical potentials. The trace extends to a subset of the Hilbert
space specified by $n$ order parameters $\Delta_a$ with $a=1,...,n$,
which are invariant under the symmetry group of ${\cal H}$ and are 
homogeneous.
In the mean field approach, one approximates ${\cal H}$ with 
${\cal H}_{\rm MF}$ that depends 
on the order parameters explicitly and leaves the trace unrestricted. 
In the superfluid of imbalanced atoms or in g2SC, we have $m=2$ 
and $n=1$. $Q_1$ and $Q_2$ stand for 
the number of atoms of each species in the former or the electric charge and the 8th color charge 
in the latter. $\Delta_1$ corresponds to the magnitude of the gap parameter in both cases. 
In particular, $\Delta_1\equiv\Delta$ of the last two sections.
For gCFL, $m=n=3$ with $Q_1$, $Q_2$ and $Q_3$ the electric, 3rd color and 8th color 
charges while $\Delta_1$, $\Delta_2$ and $\Delta_3$ correspond to the magnitude of the 
eigenvalues of the $3\times 3$ diquark condensate matrix
$\Phi_f^c\equiv \epsilon^{cc_1c_2}\epsilon_{ff_1f_2}<q_{f_1}^{c_1}q_{f_2}^{c_2}>$.
The equilibrium conditions read
\begin{equation}
\Big(\frac{\partial\Omega}{\partial\Delta_a}\Big)_{\nu_1,...,\nu_m}=0.
\label{equil1}
\end{equation}
and the expectation value of the $Q$-quanta density reads
\begin{equation}
n_j=\frac{1}{V}<Q_j>=-\Big(\frac{\partial\Omega}{\partial\mu_j}\Big)_{\Delta_1,...,\Delta_n}
\label{constraint}
\end{equation}
where $V$ is the volume of the system and the thermal average $<Q_j>$ is defined as 
\begin{equation}
<Q_j>=\frac{{\rm Tr}Q_j\exp\left(-\frac{{\cal H}-\sum_{j=1}^m\mu_jQ_j}{T}\right)}
{{\rm Tr}\exp\left(-\frac{{\cal H}+\sum_{j=1}^m\mu_jQ_j}{T}\right)}.
\end{equation}
The Sarma instability corresponds to the negative eigenvalues of the
$n\times n$ stability matrix
\begin{equation}
\Big(\frac{\partial^2\Omega}{\partial\Delta_a\partial\Delta_b}\Big)_
{\mu_1,...,\mu_n}.
\label{sarma}
\end{equation}
Under the constraints of fixed densities $n_j$'s,
the relevent thermodynamic quantity to be minimized is the Helmholtz
free energy which is obtained by a Legendre transformation,
\begin{equation}
{\cal F}=\Omega+\sum_j\mu_jn_j.
\end{equation}
While the equilibrium conditions
\begin{equation}
\Big(\frac{\partial{\cal F}}{\partial\Delta_a}\Big)_{n_1,...,n_m}=0.
\end{equation}
are equivalent to Eq.~(\ref{equil1}), the stability matrix becomes
\begin{equation}
\Big(\frac{\partial^2{\cal F}}{\partial\Delta_a\partial\Delta_b}\Big)_
{n_1,...,n_m}
=\Big(\frac{\partial^2\Omega}{\partial\Delta_a\partial\Delta_b}\Big)_
{\mu_1,...,\mu_m}
+\sum_{ij}(M^{-1})_{ij}\Big(\frac{\partial Q_i}{\partial\Delta_a}\Big)_
{\mu_1,...,\mu_m}\Big(\frac{\partial Q_j}{\partial\Delta_b}\Big)_
{\mu_1,...,\mu_m}.
\label{helmholtz}
\end{equation}
where the $m\times m$ matrix
\begin{equation}
M_{ij}=-\Big(\frac{\partial^2\Omega}{\partial\mu_i\partial\mu_j}\Big)_
{\Delta_1,...,\Delta_n}
\label{mmatrix}
\end{equation}
is positive. To see this point, we carry out the second order derivative and 
obtain that
\begin{equation}
M_{ij}=<Q_iQ_j>-<Q_i><Q_j>=<(Q_i-<Q_i>)(Q_j-<Q_j>)>.
\end{equation}
The positivity follows from the observation that $z^\dagger M z>0$ for 
an arbirary $m\times 1$ complex column matrix $z$.
Therefore, the stability matrix (\ref{helmholtz}) could be positive
even Eq.~(\ref{sarma}) is not so. This is precisely what the constraint of
fixed particle numbers in case of BP or the constraint of the charge neutrality
accomplished. Because of the relations (\ref{constraint}) and 
\begin{equation}
\mu_j=\left(\frac{\partial{\cal F}}{\partial n_j}\right)_{\Delta_1,...,
\Delta_n},
\end{equation}
we have
\begin{equation}
M_{ij}=\left(\frac{\partial n_i}{\partial\mu_j}\right)_{\Delta_1,...,
\Delta_n}
\end{equation}
and
\begin{equation}
(M^{-1})_{ij}=\left(\frac{\partial\mu_i}{\partial n_j}\right)_{\Delta_1,
...,\Delta_n}
=\left(\frac{\partial^2{\cal F}}{\partial n_i\partial n_j}\right)_{
\Delta_1,...,\Delta_n}.
\end{equation}
Therefore the positivity of $M_{ij}$ also implies that of the matrix
\begin{equation}
\left(\frac{\partial^2{\cal F}}{\partial n_i\partial n_j}\right)_{
\Delta_1,...,\Delta_n}.
\end{equation}

The inclusion of inhomogeneous variations renders the stability matrix
infinitely dimensional with additional matrix elements
\begin{equation}
\Big(\frac{\partial^2{\cal F}}{\partial\phi_a^*(\vec k)
\partial\phi_b(\vec k)}\Big)_{n_1,...,n_m}
=\Big(\frac{\partial^2\Omega}{\partial\phi_a^*(\vec k)
\partial\phi_b(\vec k)}\Big)_{\mu_1,...,\mu_m}
\label{inhomo}
\end{equation}
at $\vec k\neq 0$, without the second term of Eq.~(\ref{helmholtz}),
where we have parametrized a general variation of the order
parameters according to
\begin{equation}
\delta\Delta_a+ \frac{1}{V}\sum_{\vec k\neq 0}
\delta\phi_a(\vec k)e^{i\vec k\cdot\vec r}
\end{equation}
with $V$ the volume of the system. If the matrix elements Eq.~(\ref{inhomo})
are not singular at $\vec k=0$ in the thermodynamic limit, $V \to\infty$,
\begin{equation}
\lim_{\vec k\to 0}\lim_{V\to\infty}\Big(\frac{\partial^2\Omega}{\partial\phi_a^*(\vec k)
\partial\phi_b(\vec k)}\Big)_{\mu_1,...,\mu_m}
=\Big(\frac{\partial^2\Omega}{\partial\Delta_a\partial\Delta_b}\Big)_
{\mu_1,...,\mu_m}
\end{equation}
The Sarma instability will show up with respect to the variations of 
a sufficiently low but nonzero momentumn.

The Higgs instability can also be understood intuitively. Let us devide the
whole system being considered into ${\cal N}$ subsystems of equal volumes and equal 
chemical potentials. We label them by the integer $p$. As long as the size of each
subsystem is much larger than all microscopic length scales, the energy of the interfaces 
can be ignored. At equilibrium, the thermodynamic potential density of the master 
system is 
\begin{equation}
\Omega^{\rm tot.}=\frac{1}{{\cal N}}\sum_p\Omega^{(p)}=\Omega_M
\end{equation}
where $\Omega_M$ is given by (\ref{omegaM}) with $\Delta$ by (\ref{gapsol}). Now we consider an 
inhomogeneous variation of the gap parameters, $\delta\Delta_1^{(p)},...,\delta\Delta_n^{(p)}$,
such that 1) each of the $n$-component vector, $(\Delta_1^{(p)},...,\Delta_n^{(p)})$ 
proportional to the same eigenvector of the matrix
\begin{equation}
\left(\frac{\partial^2\Omega}{\partial\Delta_a\partial\Delta_b}\right)_{\mu_1,...,\mu_m}
\end{equation}
with the negative eigenvalue $\lambda<0$, 2) $\sum_p\delta\Delta_a^{(p)}=0$. Then
\begin{equation}
\delta n_j=\frac{1}{{\cal N}}\sum_{a,p}\frac{\partial n_j}{\partial\Delta_a}\delta\Delta_a^{(p)}
=\frac{1}{{\cal N}}\sum_a\frac{\partial n_j}{\partial\Delta_a}\sum_p\delta\Delta_a^{(p)}=0
\end{equation}
but 
\begin{equation}
\delta\Omega^{\rm tot.}\simeq\frac{1}{{\cal N}}\sum_{a,b}\frac{\partial^2\Omega}
{\partial\Delta_a\partial\Delta_b}\sum_p\delta\Delta_a^{(p)}\delta\Delta_b^{(p)}
=\frac{1}{{\cal N}}\lambda\sum_{a,p}\delta\Delta_a^{(p)}\delta\Delta_a^{(p)}<0.
\end{equation}
Therefore the Sarma instability develops through
different variations of the order parameters of each susystem while
maintaining the constraint conditions of the master system.

\section{Coulomb Energy}
\label{sec-com}

\vskip 4mm

Unless a competing mechanism that results in a positive contribution to the inhomogeneous 
blocks of the stability matrix. The Higgs instability will prevent gapless superfluidity/
superconductivity from being implemented in nature. In the system of imbalanced neutral 
atoms, such a mechanism is not likely to exist and this contributes to the reason why the 
BP state has never been observed there. For the quark matter being considered, however, the positive 
Coulomb energy induced by the Higgs field of electrically charged diquark pairs has to be 
examined. 

The second order derivative of the Hemlholtz free energy, Eq.~(\ref{helmholtz}) 
for g2SC reads: 
\begin{equation}
\Big(\frac{\partial^2{\cal F}}{\partial\Delta^2}\Big)_{n_e} 
= \Big(\frac{\partial^2\Omega}{\partial\Delta^2}\Big)_{\mu_e}
+\frac{\Big(\frac{\partial n_e}{\partial\Delta}\Big)_{\mu_e}^2}
{\Big(\frac{\partial n_e}{\partial\mu_e}\Big)_{\Delta}}
\label{free}
\end{equation}
and the stability implemented by the charge neutrality implies that 
\begin{equation}
\Big(\frac{\partial^2{\cal F}}{\partial\Delta^2}\Big)_{n_e}>0.
\end{equation}

\begin{figure}[ht]
\vspace*{-3.5truecm}
\includegraphics[width=16cm]{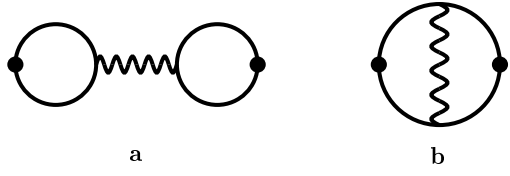}
\vspace*{-16truecm}
\caption{\label{Coulomb-f}
$a$ is the Coulomb Energy in Eq. (\ref{coulomb}), and $b$ is the ignored exchange 
Coulomb energy.}
\end{figure}

While the global charge neutrality is maintained, an inhomogeneous $\phi$ will induce local 
charge distribution. The corresponding Coulomb energy
\begin{equation}
E_{\rm coul.}=\frac{1}{2}\sum_{\vec k\neq 0}\frac{\delta\rho(\vec k)^*\delta\rho(\vec k)}
{k^2+m_D^2(k)},
\label{coulomb}
\end{equation}
should be added to the RHS of the stability matrix element (\ref{inhomo})
where $\delta\rho(\vec k)$ is the Fourier component of the $\phi$-induced charge density
and $m_D$ is the Coulomb polarization function ( Debye mass at $\vec k=0$ ). We have 
\begin{equation}
m_D^2(k)=-\frac{e^2T}{2}\sum_P{\rm tr}\gamma_0Q{\cal S}(P+K)\gamma_0Q
{\cal S}(P)
\end{equation}
and 
\begin{equation}
\delta\rho(\vec k)=\kappa(k)H(\vec k),
\end{equation}
where 
\begin{equation}
\kappa(k)=\frac{ieT}{2}\sum_P{\rm tr}\gamma_0Q{\cal S}(P+K)
\gamma_5\epsilon^3\rho_2{\cal S}(K)
\end{equation}
with $K=(0,\vec k)$ and $Q$ the electric charge operator. We have  
\begin{equation}
Q=\rho_3(a+b\tau_3)
\end{equation}
with $a=1/6$ and $b=1/2$ for the quark matter consisting of 
$u$ and $d$ flavors. 
The Eq.~(\ref{coulomb}) corresponds to 
the diagram $a$ of Fig. \ref{Coulomb-f}. The diagram $b$ in Fig. \ref{Coulomb-f} 
represents the exchange Coulomb energy and is 
ignored since the typical momentum running through the Coulomb line of the order $\bar\mu$ 
compared that in the first diagram, $m_D\sim \alpha_e\bar\mu$. It is shown in appendix C that 
the Goldstone fields do not generate electric charge distributions.

The inhomogeneous block of the stability matrix element, (\ref{inhomo}), becomes now
\begin{eqnarray}
\tilde\Pi_H(k) &\equiv&
\Big(\frac{\partial^2{\cal F}}{\partial H^*(\vec k)\partial H(\vec k)}\Big)_{n_e}
=\Big(\frac{\partial^2\Omega}{\partial H^*(\vec k)\partial H(\vec k)}\Big)_{\mu_e}
+\frac{\kappa^*(k)\kappa(k)}{k^2+m_D^2(k)}\\ \nonumber
&=&\Pi_{H}(k)+\frac{\kappa^*(k)\kappa(k)}{k^2+m_D^2(k)}.
\label{higgscoul}
\end{eqnarray}
The stability of the system with respect to the Higgs field requires that 
$\tilde\Pi_H(k)>0$ for all $k$.

The form factor $\kappa(q)$ and the momentum dependent Debye mass square 
can be calculated explicitly at weak coupling, i.e.
$\delta\mu<<\bar\mu$ and $k<<\bar\mu$. We find that 
\begin{equation}
\kappa(k)=\frac{2e^2\bar\mu^2b}{\pi}K(k|\delta\mu)
\end{equation}
with 
\begin{equation}
K(k|\delta\mu)=i\Delta T\sum_n\frac{\omega_n+i\delta\mu}
{\sqrt{(\omega_n+i\delta\mu)^2+\Delta^2}}\int_{-1}^1dx
\frac{1}{(\omega_n+i\delta\mu)^2+\Delta^2+\frac{1}{4}k^2x^2}
\label{Kintegral}
\end{equation}
and 
\begin{equation}
m_D^2(k)=\frac{6(a^2+b^2)e^2\bar\mu^2}{\pi^2}-\frac{2b^2e^2\bar\mu^2}{\pi}I(k|\delta\mu)
\end{equation}
with the first term the Debye mass of the normal phase and $I(q|\delta\mu)$ the function 
defined in the section III. 

\begin{figure}
\includegraphics[width=10cm]{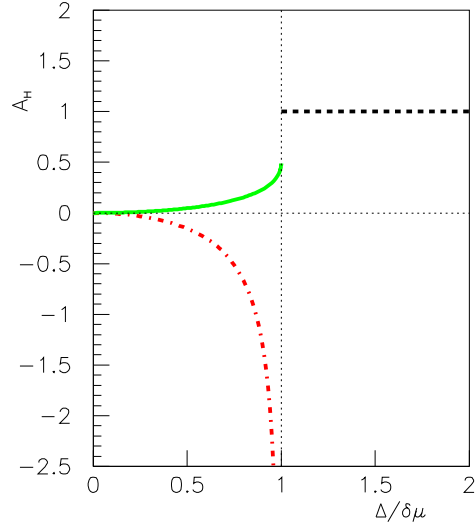}
\caption{\label{Higgs-f-fig}
$A_H$(red dashed-dotted line) and $\tilde A_H$ (green solid line)
are plotted as functions of $\Delta/\delta\mu$ in the g2SC phase, and the
black dashed line shows the corresponding quantity in the 2SC phase.}
\end{figure}

\vskip 4mm
{\bf  Zero momentum limit:} 
\vskip 4mm

Notice that 
\begin{equation}
\lim_{k\to 0}\lim_{V\to\infty}\Big(\frac{\partial^2\Omega}{\partial H^*(\vec k)\partial H(\vec k)}\Big)_{\mu_Q}
= \Big(\frac{\partial^2\Omega}{\partial\Delta^2}\Big)_{\mu_Q},
\end{equation}
\begin{equation}
m_D^2(0)=e^2\Big(\frac{\partial n_e}{\partial\mu_e}\Big)_{\Delta,\mu_B}
\end{equation}
and
\begin{equation}
\kappa(0)=e\Big(\frac{\partial n_e}{\partial\Delta}\Big)_{\mu_e,\mu_B}.
\end{equation}
We have 
\begin{equation}
\lim_{\vec k\to 0}\Big(\frac{\partial^2{\cal F}}{\partial H^*(\vec k)\partial H(\vec k)}\Big)_{\mu_e}
= \Big(\frac{\partial^2{\cal F}}{\partial\Delta^2}\Big)_{\mu_e},
\end{equation}
and the charge neutrality stabilize also the inhomogeneous Higgs field with the momentum 
much smaller than the inverse coherence length and the inverse Debye length.

\begin{figure}
\includegraphics[width=10cm]{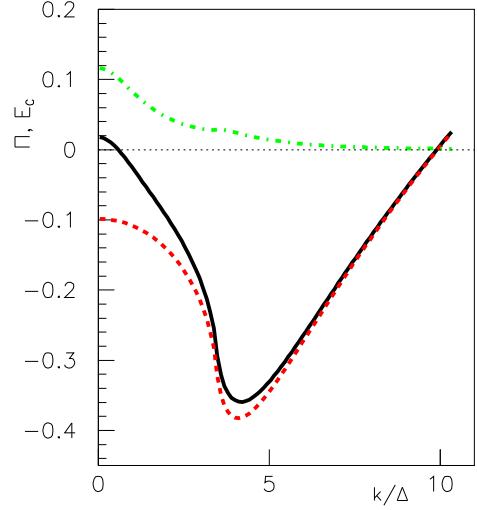}
\caption{\label{Higgs-k-fig}
The function $\Pi_H(k)$ (red dashed line), $\tilde\Pi_H(k)$ (black solid line) and the Coulomb
energy (green dash-dotted line) as functions of scaled-momentum $k/\Delta$,  in the
case of $\delta\mu=2\Delta$ and $(e^2\bar\mu^2)/(4\pi\Delta^2)=1$.}
\end{figure}

In the static long-wave length limit, we have
\begin{equation}
m_D^2(0)=\frac{2e^2b^2\bar\mu^2}{\pi^2}\Big(1+\frac{2\delta\mu}
{\sqrt{\delta\mu^2-\Delta^2}}\Big)
\label{debye0}
\end{equation}
and
\begin{equation}
\kappa(0)=\frac{4eb\bar\mu^2}{\pi^2}\frac{\Delta}{\sqrt{\delta\mu^2-\Delta^2}}.
\label{susc0}
\end{equation}
The positivity of the Debye mass square is ensured by the positivity of the $M$-
matrix defined in (\ref{mmatrix}). 
It follows from Eqs. (\ref{sarmains}), (\ref{higgscoul}), (\ref{debye0}) and 
(\ref{susc0}) that the Higgs self-energy including Coulomb energy correction
\begin{equation}
{\tilde A}_H \equiv \tilde\Pi_{H}(0) =
\Big(\frac{\partial^2{\cal F}}{\partial\Delta^2}\Big)_{\mu,n_Q}
=\frac{4(b^2-3a^2)\bar\mu^2
(\delta\mu-\sqrt{\delta\mu^2-\Delta^2})}{\pi^2[3a^2\sqrt{\delta\mu^2-\Delta^2}
+b^2(2\delta\mu+\sqrt{\delta\mu^2-\Delta^2})]}
\end{equation}
is always positive for the whole range of g2SC state. It means that the
Sarma instability in the gapless phase can be cured by Coulomb energy.
This is shown in Fig.~\ref{Higgs-f-fig}, where the red dash-dotted line indicates
the $A_H$ and the green solid line indicates ${\tilde A}_H$. 

\vskip 4mm
{\bf  Nonzero momentum case: }
\vskip 4mm
  
While the Sarma instability in g2SC phase can be cured by
Coulomb energy under the constraint of charge neutrality condition,
it is {\it{not sufficient}} for the system to be stable, even if the chromomagnetic 
instabilities are removed, say by gluon condensation. One has to explore
the Higgs instability by calculating the self-energy function 
$\tilde\Pi(k)$ in the whole momentum space, which amounts to value the three basic functions
$I(k|\delta\mu)$, $J(k|\delta\mu)$ and $K(k|\delta\mu)$ defined in (\ref{Iintegral}), 
(\ref{Jintegral}) and (\ref{Kintegral}). At $T=0$, the summation over Matsubara frequencies 
becomes an integral and we have
\begin{equation}
I(k|\delta\mu)=\frac{\Delta^2}{2\pi}\int_{-1}^1dx\Big(
\int_{-\infty+i\delta\mu}^{i\delta\mu-0^+}+\int_{i\delta\mu+0^+}^{\infty+i\delta\mu}
\Big)dz\frac{1}{\sqrt{z^2+\Delta^2}\Big(z^2+\Delta^2+\frac{1}{4}k^2x^2\Big)}
\end{equation}
\begin{equation}
J(k|\delta\mu)=\frac{\Delta^2}{2\pi}\int_{-1}^1dx x^2\Big(
\int_{-\infty+i\delta\mu}^{i\delta\mu-0^+}+\int_{i\delta\mu+0^+}^{\infty+i\delta\mu}\Big)
dz\frac{1}{\sqrt{z^2+\Delta^2}\Big(z^2+\Delta^2+\frac{1}{4}k^2x^2\Big)}
\end{equation}
and
\begin{equation}
K(k|\delta\mu)=\frac{i\Delta}{2\pi}\int_{-1}^1dx\Big(
\int_{-\infty+i\delta\mu}^{i\delta\mu-0^+}+\int_{i\delta\mu+0^+}^{\infty+i\delta\mu}\Big)
dz\frac{z}{\sqrt{z^2+\Delta^2}\Big(z^2+\Delta^2+\frac{1}{4}k^2x^2\Big)}.
\end{equation}
Further reductions of these integrals are straightforward but rather technical. The details are 
deferred to Appendix \ref{k-space}. 
Here we present only the numerical results in Fig. \ref{Higgs-k-fig} and Table I. 

Fig.~\ref{Higgs-k-fig} shows the Higgs self-energy $\Pi_H(k)$ (the red dashed line)
the Coulomb corrected Higgs self-energy ${\tilde \Pi}_H(k)$ (the black solid line)
and the Coulomb energy $E_{coul}$ (the green dash-dotted line) as functions of 
scaled-momentum $k/\Delta$,  in the case of $\delta\mu=2\Delta$ and 
$(e^2\bar\mu^2)/(4\pi\Delta^2)=1$.

Eventhough the Higgs instability can be removed by the Coulomb energy for small momenta, it returns 
for intermediate momenta. This phenomenon persists for a wide range of gap magnitude,
$0<\Delta<0.866\delta\mu$ and for all strength of the Coulomb interaction, measured by 
the dimensionless ration $\eta\equiv\frac{\alpha_e\bar\mu^2}{\Delta^2}$. Within 
the narrow range $0.866\delta\mu<\Delta<\delta\mu$, the Higgs instability could be 
removed if the Coulomb interaction were sufficiently strong. For $\eta=1$, We found that 
$\tilde\Pi_H(k)>0$ for all $k$ if $0.998\delta\mu<\Delta<\delta\mu$. In terms of the 
values of the parameters of NJL model, we have $\alpha_e\bar\mu^2<\delta\mu$. Therefore 
the electric Coulomb energy cannot cure the Higgs instability for a realistic two 
flavor quark matter.  

Negative $\Pi_H(k)$ indicates the Higgs mode is unstable 
and will decay \cite{Coleman-Weinberg}.
It is noticed that $\Pi_H(k)$ reaches its minimum at a momentum, 
i.e., $k \simeq 4 \Delta$, which indicates that a stable state may develop
around this minimum, we characterize this momentum as $k_{min}$. 
The inverse $k_{min}^{-1}$ is the typical wavelength for the unstable 
mode \cite{Weinberg-Wu}.
If mixed phase can be formed, the typical size $l$ of the 2SC bubbles 
should be as great as $k_{min}^{-1}$, i.e., $l \simeq k_{min}^{-1}$ 
\cite{Weinberg-Wu}, which turns out to becomparable to the coherence length of 2SC
in accordance with Eq.~(\ref{gapsol}).
(In the case of  homogeneous superconducting phase, $k_{min}=0$, 
and $l\rightarrow \infty$.)
Considering that the coherence length $\xi$ of a superconductor is proportianl to 
the inverse of the gap magnitude, i.e., $\xi \simeq \Delta^{-1}$ \cite{Type-trans},
therefore, a rather large ratio of $k_{min}/\Delta$ means
a rather small ratio of $l/\xi$. When $l/\xi <1$, a phase separation state is 
more favorable \cite{Igor}.

\begin{table}
\begin{tabular}{c|r|r|r|r|r}
$\delta\mu/\Delta$\kern8pt& $\alpha_e\Delta^2/\bar\mu^2$\kern8pt& $\tilde\Pi(0)$ 
\kern8pt& $\tilde\Pi_{\rm min}(k)$ \kern8pt& $k_{\rm min}/\Delta$ \kern8pt& $k_{\rm min}/\Delta_0$\kern8pt\\
\hline
5 \kern8pt& 0.25\kern8pt& 0.0026\kern8pt& -0.2778\kern8pt& 11.72\kern8pt& 1.18\kern8pt\\
\hline
5 \kern8pt& 1\kern8pt& 0.0026\kern8pt& -0.2774\kern8pt& 11.71\kern8pt& 1.18\kern8pt\\
\hline
5\kern8pt& 4\kern8pt& 0.0026\kern8pt& -0.2763\kern8pt& 11.71\kern8pt& 1.18\kern8pt\\
\hline
2\kern8pt& 0.25\kern8pt& 0.0180\kern8pt& -0.3760 \kern8pt& 4.12\kern8pt& 1.10\kern8pt\\
\hline
2\kern8pt& 1\kern8pt& 0.0180\kern8pt& -0.3592\kern8pt& 4.12\kern8pt& 1.10\kern8pt\\
\hline
2\kern8pt& 4\kern8pt& 0.0180\kern8pt& -0.3148\kern8pt& 4.12\kern8pt& 1.10\kern8pt\\
\hline
1.25\kern8pt& 0.25\kern8pt& 0.0606\kern8pt& -0.6377\kern8pt& 1.78\kern8pt& 0.89\kern8pt\\
\hline
1.25\kern8pt& 1\kern8pt& 0.0606\kern8pt& -0.4185\kern8pt& 1.78\kern8pt& 0.89\kern8pt\\
\hline
1.25\kern8pt& 4\kern8pt& 0.0606\kern8pt& -0.2045\kern8pt& 1.78\kern8pt& 0.89\kern8pt\\
\hline
\end{tabular}
\bigskip
\caption{The minimum of the self-energy of Higgs field for various 
mismatch parameters and strengths of Coulomb interaction.}
\label{table-I}
\end{table}

In Table \ref{table-I}, we show the Coulomb corrected Higgs self-energy 
at zero momentum $\tilde{\Pi}_H(0)$, its minimum, the corresponding
$k_{min}/\Delta$ and $k_{\rm min}/\Delta_0$ ( with $\Delta$ and $\Delta_0$ related by eq.
(\ref{gapsol}) for different parameters of $\Delta/\delta\mu$ and 
$\alpha_e\Delta^2/\bar\mu^2$. It is found that the momentum for
the Higgs self-energy to reach minimun is not sensitive
to the electromagnetic coupling strength, rather very sensitive to the ratio
of $\delta\mu/\Delta$. Larger $\delta\mu/\Delta$ is, larger $k_{min}/\Delta$
will be. But the ratio of this momentum to the gap of 2SC under the same pairing strength
is close to one for all cases being examined. 

The reader has to keep in mind that our result of the Higgs instability only indicates 
some kind of inhomogeneous states whose typical length scale is comparable to the 
coherence length of 2SC. Further insight on the structure of the inhomogeneity 
cannot be gained without exploring the higher order terms of the nonlinear 
realization (\ref{II}). A more direct approach to obtain the favorite structure of 
the ground state is to compare the free energy of various candidate states, which 
include the mixed phase, the single-plane wave FF state, striped LO state and
multi-plane wave states. The Coulomb energy and the gradient energy have to be 
estimated reliably.  We leave this analysis as a future project.

\section{Summary and overlook}
\label{sec-summary}

To understand the pair-breaking superfluidity or superconductivity induced by 
mismatched Fermi momenta, one has to go beyond the standard framework of
BCS theory. In this paper, we offered a nonlinear realization framework of the 
gauged Nambu--Jona-Lasinio model, where the fluctuations of the Higgs and 
pseudo Nambu-Goldstone fields can be investigated simutaneously. 
 
It is found that, with the increase of mismatch between the Fermi surfaces of the 
pairing quarks, not only Nambu-Goldstone currents can be spontaneously 
generated, but the Higgs filed becomes unstable in the gapless phase. 
As we already knew, the instability of the Nambu-Goldstone currents naturally 
induce the formation of the single-plane wave FF state. 
In this paper, we focused on the Higgs instability which has not drawn
much attention yet. 

The Higgs instability is the origin of spatial inhomogeneity, and is not 
prohibited by global constraints.
In the case of the imbalanced neutral atom systems, the instability
extends from arbitrarily low momenta-albeit not zero-to momenta
much higher than the inverse coherence length. Without a long range force 
a macroscopic phase separation is likely to be the structure of the ground state since 
it minimizes the gradient energy of the inhomogeneity without additional costs.
The thermodynamics of the phase separation can be obtained following the Gibbs 
construction of \cite{gibbs}. 

For the case of two flavor quark matter that is globally neutral, the instability
is absent for high momenta and 
can be removed by the electric Coulomb energy for low momenta.
But it remains in a window of intermediate momenta. In the regime where
$\delta\mu >> \Delta$, the wavelength for the 
unstable Higgs mode is alway comparable with the gap magnitude of 2SC phase, which indicates 
the tendency to form 2SC bubbles. Because of the Long range Coulomb force, a macroscopic 
phase separation is prohibited while a crystalline structure is more favorable. 
Possible candidates
include the heterotic mixture of normal and superconducting phases
and the multi-plane-wave LOFF state. 

Although our calculations are
limited to the case of g2SC, the existence of the Higgs instability
in the absence of long range  interactions is generic for all
gapless superfluidity/superconductivity that are subject to the
Sarma instability. Our general formulation can be readily applied 
to the system with several invariant order parameters and constraints. It 
would be interesting to examine whether the electric and color
Coulomb energies are capable of eliminating the Higgs instability completely
in the gCFL phase.
 
It is quite surprising that the Higgs instability shows up at some finite momentum, though 
it is removed by the constraint of the charge neutrality. 
Following the discovery of the chromomagnetic instability of g2SC at $T=0$, the authors 
of \cite{wangqh} examined the Meissner mass square at $T\neq 0$ and found that the 
instability goes away for $T$ sufficiently close to $T_c$. In the light of our result 
of the Higgs instability, this criterion is not sufficient for the stability. One has to extend the 
analysis of \cite{wangqh} to all momenta of the chromomagnetic polarization function as well 
as that of the Higgs self-energy function  at $T\neq 0$ in order to map out the 
stability domain of g2SC in the phase diagram of the two flavor quark matter. This amounts 
to value the functions $I(k|\delta\mu)$, $J(k|\delta\mu)$ and $K(k|\delta\mu)$ at nonzero 
$T$, which is straightforward. We hope to report our progress in this direction in near 
future.


\section*{Acknowledgments}
We thank M. Forbes, K. Fukushima, K. Iida, M. Hashimoto, T. Hatsuda, D.K. Hong, 
I. Shovokovy and  P.F. Zhuang for stimulating discussions. The work of I.G. and H.C.R 
is supported in part by US Department of Energy 
under grants DE-FG02-91ER40651-TASKB. The work of D.F.H. is supported in part 
by Educational Committee under grants NCET-05-0675 and 704035. The work of D.F.H 
and H.C.R. is also supported in part by NSFC under grant No. 10575043.  The work of 
M.H. is supported in part by the Institute of High Energy Physics, Chinese Academy of 
Sciences and the Japan Society for the Promotion of Science Fellowship 
Program, and she also would like to thank APCTP and KEK for the hospitality.

\appendix

\section{Calculation of $\Pi_H(k)$ and $\tilde\Pi_H(k)$}
\label{Higgs-bubble}

The calculation of the one-loop self energies of the inhomogeneously fluctuating 
fileds introduced in Section \ref{sec-fluc} and \ref{sec-com} will be illurstrated in 
this appendix with that of the Higgs field, given by
\begin{equation}
\Pi_H(k)=\frac{1}{2G_D}-\frac{1}{2}T\sum_n\int\frac{d^3\vec p}{(2\pi)^3}
{\rm Tr}\epsilon^3\rho_2\tau_2{\cal S}_M(\omega_n,\vec p+\vec k)
\epsilon^3\rho_2\tau_2{\cal S}_M(\omega_n,\vec p),
\end{equation}
where we exihibite the dependence of the propagator on momentum and Matsubara 
energy separately. 
It is convenient to tranform the propagator into the representation of Ref.~\cite{FF-Ren} via 
\begin{equation}
{\cal S}_M(\omega,\vec p)\equiv U{\cal S}(\omega,\vec p)U^\dagger
\end{equation}
where $U=\rho_++\rho_-\tau_2$ with $\rho_\pm=\frac{1}{2}(1\pm\rho_3)$. 
It follows that
\begin{equation}
{\cal S}(\omega,\vec p)^{-1}
=i\omega\gamma_0-\vec\gamma\cdot\vec p+\bar\mu\gamma_0\rho_3-\delta\mu\gamma_0\tau_3
+\Delta\gamma_5\epsilon^3\rho_2,
\label{newrep}
\end{equation}
where we have suppressed the chemical potential associated to the 8th color.
Correspondingly, 
\begin{equation}
\Pi_H(k)=\frac{1}{2G_D}-\frac{1}{2}T\sum_n\int\frac{d^3\vec p}{(2\pi)^3}
{\rm Tr}\epsilon^3\rho_2{\cal S}(\omega_n,\vec p+\vec k)
\epsilon^3\rho_2{\cal S}(\omega_n,\vec p),
\end{equation}
where the matrix under the trace is diagonal in isospin space and the trace 
with respect to iospins becomes the sum over the eigenvalue of $\tau_3$.

With respect to the color indexes r, g and b,
\begin{equation}
\epsilon^3=\left(\begin{array}{cc}
i\sigma_2 & 0 \\
0 & 0 \end{array}\right)
\end{equation}
with $\sigma_2$ the second Pauli matrix in the color SU(2) subspace. 
Accordingly, 
\begin{equation}
\Pi_H(k)=\frac{1}{2G_D}-\frac{1}{2}T\sum_n\int\frac{d^3\vec p}{(2\pi)^3}
{\rm Tr}\epsilon^3\rho_2{\cal S}^{SU(2)}(\omega_n,\vec p+\vec k)
\epsilon^3\rho_2{\cal S}^{SU(2)}(\omega_n,\vec p),
\label{higgsform}
\end{equation}
with ${\cal S}^{SU(2)}$ the color SU(2) block of the propagator. The expression 
of ${\cal S}^{SU(2)}$ can be obtained from (\ref{newrep}) with the replacement 
of $\epsilon^3$ by $i\sigma_2$. The trace in color space extends only to r and g 
indexes now. 

On writing  
\begin{equation}
[{\cal S}^{SU(2)}(\omega,\vec p)]^{-1}=\gamma_0\rho_3[\rho_3(i\omega-\vec\alpha\cdot\vec p)
+\bar\mu-\delta\mu\rho_3\tau_3+i\Delta\gamma_0\gamma_5\sigma_2\rho_1],
\end{equation}
we notice that the matrix (...) can be projected into two orthogonal subspace via
\begin{equation}
\Lambda_{\pm}=\frac{1}{2}(1\pm\rho_3\vec\alpha\cdot\hat p)
\end{equation}
with $\vec\alpha=\gamma_0\vec\gamma$. We find that
\begin{eqnarray}
&{}&(i\omega-\vec\alpha\cdot\vec p)\rho_3+\bar\mu-\delta\mu\rho_3\tau_3
+i\Delta\gamma_0\gamma_5\sigma_2\rho_1
\nonumber \\
&=& \Lambda_+(i\omega\rho_3-p+\bar\mu-\delta\mu\rho_3\tau_3
+i\Delta\gamma_0\gamma_5\sigma_2\rho_1)
+\Lambda_-(i\omega\rho_3+p+\bar\mu-\delta\mu\rho_3\tau_3
+i\Delta\gamma_0\gamma_5\sigma_2\rho_1).
\end{eqnarray}
Under the weak coupling approximation, $\omega<<\bar\mu$, $|p-\bar\mu|<<\bar\mu$, 
$\Delta<<\bar\mu$ and $\delta\mu<<\bar\mu$,
The contribution of the second term to the inverse can be ignored. We obtains that
\begin{equation}
{\cal S}_M^{SU(2)}(\omega,\vec p)\simeq \Lambda_+
(i\omega\rho_3-p+\bar\mu-\delta\mu\rho_3\tau_3+i\Delta\gamma_0\gamma_5\sigma_2\rho_1)^{-1}
=-\frac{\gamma_0\rho_3-\vec\gamma\cdot\hat p}{2}
\frac{(i\omega-\delta\mu\tau_3)\rho_3+p-\bar\mu+\Delta\gamma_0\gamma_5\sigma_2\rho_1}
{(\omega+i\delta\mu\tau_3)^2+(p-\bar\mu)^2+\Delta^2}.
\label{weak}
\end{equation}

Substituting Eq.~(\ref{weak}) into Eq.~(\ref{higgsform}) and carried out the trace over 
Dirac indexes, NG indexes and SU(2) color indexes, we end up with  
\begin{eqnarray}
\Pi_H(k)&=&\frac{1}{2G_D}-4T\sum_{n,\tau_3=\pm}\int\frac{d^3\vec p}{(2\pi)^3}
\frac{(\omega_n+i\delta\mu\tau_3)^2+\xi^\prime\xi-\Delta^2}
{[(\omega_n+i\delta\mu\tau_3)^2+\xi^{\prime 2}+\Delta^2][(\omega_n+i\delta\mu\tau_3)^2+\xi^2+\Delta^2]}
\nonumber \\
&=&\frac{1}{2G_D}-8T\sum_n\int\frac{d^3\vec p}{(2\pi)^3}
\frac{(\omega_n+i\delta\mu)^2+\xi^\prime\xi-\Delta^2}
{[(\omega_n+i\delta\mu)^2+\xi^{\prime 2}+\Delta^2][(\omega_n+i\delta\mu)^2+\xi^2+\Delta^2]}
\label{after}
\end{eqnarray}
where $\xi=p-\bar\mu$ and $\xi^\prime=p^\prime-\bar\mu$ with 
$\vec p^\prime = \vec p+\vec k$. The 2nd equality follows from relabeling 
the index of the sum over $n$ such that  $\omega_n\to -\omega_n$. In deriving 
(\ref{after}) we employed the approximation $\hat p\simeq\hat p^\prime$.
The integral over $\vec p$ can be approximated according to
\begin{equation}
\int d^3\vec p\simeq 2\pi\bar\mu\int_{-1}^1d\cos\theta\int_{-\infty}^\infty d\xi
\end{equation}
and the $\xi$-integration can be carried out with residues ( notice that 
$\xi^\prime-\xi\simeq k\cos\theta$ ). We have
\begin{equation}
\Pi_H(k)=\frac{1}{2G_D}-\frac{2\bar\mu^2T}{\pi}\sum_n\frac{1}{\sqrt{(\omega_n+i\delta\mu)^2
+\Delta^2}}\int_{-1}^1dx\frac{(\omega_n+i\delta\mu)^2}{(\omega_n+i\delta\mu)^2
+\Delta^2+\frac{1}{4}k^2x^2}
\end{equation}
where we have denote $x=\cos\theta$ for the integration variable.
Substituting in the gap equation
\begin{equation}
\frac{1}{2G_D}=\frac{8\bar\mu^2T}{\pi}\sum_n\frac{1}{\sqrt{(\omega_n+i\delta\mu)^2
+\Delta^2}}=\frac{4\bar\mu^2T}{\pi}\sum_n\frac{1}{\sqrt{(\omega_n+i\delta\mu)^2
+\Delta^2}}\int_{-1}^1dx 1
\end{equation}
we obtain that
\begin{equation}
\Pi_H(k)=\frac{2\bar\mu^2T}{\pi}\sum_n\frac{1}{\sqrt{(\omega_n+i\delta\mu)^2
+\Delta^2}}\int_{-1}^1dx\frac{\Delta^2+\frac{1}{4}q^2x^2}{(\omega_n+i\delta\mu)^2
+\Delta^2+\frac{1}{4}q^2x^2}.
\end{equation}
and this derives Eq.~(\ref{higgsenergy}). 

\section{Calculation of integrals}
\label{k-space}

In this appendix, we shall calculate the functions $I(k|\delta\mu)$, 
$J(k|\delta\mu)$ and $K(k|\delta\mu)$ at $T=0$ and for $\delta\mu>\Delta$ We have
\begin{equation}
I(k|\delta\mu)=\frac{\Delta^2}{2\pi}\int_{-1}^1dx\Big(
\int_{-\infty+i\delta\mu}^{i\delta\mu-0^+}+\int_{i\delta\mu+0^+}^{\infty+i\delta\mu}
\Big)dz\frac{1}{\sqrt{z^2+\Delta^2}\Big(z^2+\Delta^2+\frac{1}{4}k^2x^2\Big)}
\end{equation}
\begin{equation}
J(k|\delta\mu)=\frac{\Delta^2}{2\pi}\int_{-1}^1dx x^2\Big(
\int_{-\infty+i\delta\mu}^{i\delta\mu-0^+}+\int_{i\delta\mu+0^+}^{\infty+i\delta\mu}\Big)
dz\frac{1}{\sqrt{z^2+\Delta^2}\Big(z^2+\Delta^2+\frac{1}{4}k^2x^2\Big)}
\end{equation}
and
\begin{equation}
K(k|\delta\mu)=\frac{i\Delta}{2\pi}\int_{-1}^1dx\Big(
\int_{-\infty+i\delta\mu}^{i\delta\mu-0^+}+\int_{i\delta\mu+0^+}^{\infty+i\delta\mu}\Big)
dz\frac{z}{\sqrt{z^2+\Delta^2}\Big(z^2+\Delta^2+\frac{1}{4}k^2x^2\Big)}.
\end{equation}
Because of the similarity of these integrals, we shall focus on $I(k|\delta\mu)$ below.

Notice that the $z$-integral is interupted by the branch cut from $i\Delta$ to 
infinity along the imaginary axis ( the other runs from $-i\Delta$ to infinity )
and ${\rm Re}\sqrt{z^2+\Delta^2}\ge 0$ throughout the cut $z$-plane.

\begin{figure}[ht]
\includegraphics[width=12cm]{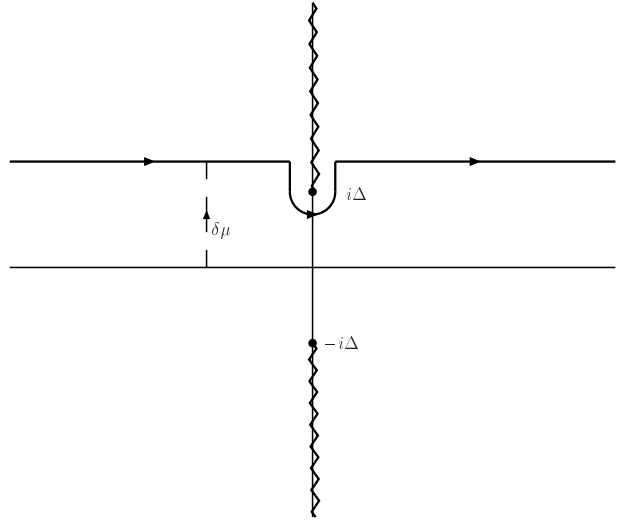}
\vspace*{-8truecm}
\caption{\label{contour-f}
The contour for $z$-integral.}
\end{figure}
Alternatively, the $z$-integral can be regarded an integral along a contour running 
from $-\infty+i\delta$ to $\infty+i\delta$ and going around the branch point 
$i\Delta$ counterclockwisely plus an integral from $i\delta\mu+0^+$ to $i\delta\mu-0^+$
going around the branch point clockwisely. The contour of the former can be shifted 
to the real axis, leaving the result independent of $\delta\mu$. We obtain that
\begin{equation}
I(k|\delta\mu)=I(k|0)+I_1(k|\delta\mu)
\end{equation}
with 
\begin{equation}
I(k|0)=\frac{\Delta^2}{2\pi}\int_{-1}^1 dx\int_{-\infty}^\infty dz
\frac{1}{\sqrt{z^2+\Delta^2}\Big(z^2+\Delta^2+\frac{1}{4}k^2x^2\Big)}
\end{equation}
and
\begin{equation}
I_1(k|\delta\mu)=\frac{\Delta^2}{2\pi}\int_{-1}^1 dx
\int_{i\delta\mu+0^+}^{i\delta\mu-0^+} dz
\frac{1}{\sqrt{z^2+\Delta^2}\Big(z^2+\Delta^2+\frac{1}{4}k^2x^2\Big)}.
\end{equation}
Using the formula
\begin{equation}
\int d\xi\frac{1}{\sqrt{\xi^2+a^2}(\xi^2+b^2)}
=\frac{1}{2b\sqrt{b^2-a^2}}\ln\frac{b\sqrt{\xi^2+a^2}+\sqrt{b^2-a^2}\xi}
{b\sqrt{\xi^2+a^2}-\sqrt{b^2-a^2}\xi}
\end{equation}
and making the transformation for the $x$-integral, $x=\frac{2\Delta}{k}
\sinh\theta$, we end up with
\begin{eqnarray}
I(k|0) &=& \frac{4\Delta}{\pi k}\int_0^{\theta_c}d\theta\frac{\theta}{\sinh\theta}
\\ \nonumber
&=& \frac{8\Delta}{\pi k}\Big[\frac{\pi^2}{8}+\frac{1}{2}\theta_c\ln
\tanh\frac{\theta_c}{2}-{\rm Li}_2(e^{-\theta_c})
+\frac{1}{4}{\rm Li}_2(e^{-2\theta_c})\Big]
\end{eqnarray}
and 
\begin{equation}
I_1(k|\delta\mu)=\frac{2\Delta}{\pi k}\int_0^{\theta_c}\frac{d\theta}{\sinh\theta}
\ln\frac{|\sqrt{\delta\mu^2-\Delta^2}\cosh\theta-\delta\mu\sinh\theta|}
{\sqrt{\delta\mu^2-\Delta^2}\cosh\theta+\delta\mu\sinh\theta},
\end{equation}
where $\theta_c$ is defined by $\sinh\theta_c=\frac{k}{2\Delta}$ and ${\rm Li}_2(z)$ 
is Spence function, i.e.
\begin{equation}
{\rm Li}_2(z)=-\int_0^z dt\frac{\ln(1-t)}{t}=\sum_{n=1}^{\infty}\frac{z^n}{n^2}.
\end{equation}

The integration of $J(k|\delta\mu)$ can be carried following the same procedure and 
the result is elementary, 
\begin{eqnarray}
J(k|\delta\mu) &=& \frac{16\Delta^3}{\pi k^3}(\theta_c\cosh\theta_c-\sinh\theta_c)
\\ \nonumber
&+& \frac{8\Delta^3}{\pi q^3}\Big[\cosh\theta_c
\ln\frac{|\sqrt{\delta\mu^2-\Delta^2}\cosh\theta_c-\delta\mu\sinh\theta_c|}
{\sqrt{\delta\mu^2-\Delta^2}\cosh\theta_c+\delta\mu\sinh\theta_c|}
-\frac{\delta\mu}{\Delta}\ln\frac{|\sqrt{\delta\mu^2-\Delta^2}-\Delta\sinh\theta_c|}
{\sqrt{\delta\mu^2-\Delta^2}+\Delta\sinh\theta_c}\Big].
\end{eqnarray}

Upon decomposing $K(k|\delta\mu)$ into $K(k|0)+K_1(k|\delta\mu)$ in analog to 
$I(k|\delta\mu)$ we observe that $K(k|0)=0$ and the integration of $K_1(k|\delta\mu)$
yields
\begin{equation}
K_1(k|\delta\mu)=\frac{4\Delta}{\pi k}{\rm Re}
\Big[{\rm Li}_2\Big(\frac{k}{2\sqrt{\delta\mu^2-\Delta^2}}\Big)
-\frac{1}{4}{\rm Li}_2\Big(\frac{k^2}{4(\delta\mu^2-\Delta^2)}\Big)\Big].
\end{equation}

The small $k$ expansion of $I(k|\delta\mu)$, $J(k|\delta\mu)$ and $K(k|\delta\mu)$ can be obtainede 
easily from their integral representations as is shown in the text. The 
large $k$ behavior of $I(k|0)$ and $J(k|\delta\mu)$ cab be extracted readily from their explicit form 
with $e^{\theta_c}>>1$. As to $I_1(k|\delta\mu)$, we write
\begin{equation}
\int_0^{\theta_c}d\theta(…)=\int_0^\infty d\theta(…)-\int_{\theta_c}^\infty d\theta(…).
\label{formula}
\end{equation}
The large $k$ expansion can be carried out with the aid of the formula 
( the derivation is outlined at the 
end of this appendix ).
\begin{equation}
\int_0^\infty\frac{d\theta}{\sinh\theta}\ln\frac{\sqrt{\delta\mu^2-\Delta^2}
\cosh\theta+\delta\mu\sinh\theta}{|\sqrt{\delta\mu^2-\Delta^2}\cosh\theta
-\delta\mu\sinh\theta|}=\frac{\pi^2}{2}
\end{equation}
and the expansion of the integrand of the 2nd integral on RHS of (\ref{formula}) according to the powers 
of $e^{-\theta}$. The large $k$ expansion of 
$K(k|\delta\mu)$ follows from the formula
\begin{equation}
{\rm Re}{\rm Li}_2(z)=\frac{\pi^2}{3}-\frac{1}{2}\ln^2z-{\rm Li}_2\Big(\frac{1}{z}\Big)
\end{equation}
for $z>1$. We obtain that
\begin{equation}
I(k|\delta\mu)\simeq\frac{4}{\pi k^2}\Big(-2\ln\frac{k}{\Delta}-2+\ln\frac{\delta\mu+\sqrt{\delta\mu^2-\Delta^2}}
{\delta\mu-\sqrt{\delta\mu^2-\Delta^2}}\Big),
\end{equation}
\begin{equation}
J(k|\delta\mu)\simeq\frac{4}{\pi k^2}\Big(2\ln\frac{k}{2}-2
+\ln\frac{\delta\mu+\sqrt{\delta\mu^2-\Delta^2}}{\delta\mu-\sqrt{\delta\mu^2-\Delta^2}}\Big)
\end{equation}
and
\begin{equation}
K(k|\delta\mu)\simeq\frac{\pi\Delta}{k}.
\end{equation}

Finally, let us sketch how the formula (\ref{formula}) can be derived. Introduce a function of $z$, 
\begin{equation}
f(z)=\int_0^\infty\frac{d\theta}{\sinh\theta}\ln\frac{1+z\tanh\theta}{1-z\tanh\theta}
\end{equation}
the LHS of (\ref{formula}) equals the real part of its value at 
$z=\frac{\delta\mu}{\sqrt{\delta\mu^2-\Delta^2}}$. For $z<1$, the integrand 
can be expanded according 
to the powers of $z$ and the integration of the coefficients can be carried 
out explicitly. The resultant 
power series can be recognized to be that of $\pi\sin^{-1}z$. The analytic 
continuation to $z>1$ gives rise to the RHS of (\ref{formula}).

\section{Electric charge density induced by Goldstone fields}

On writing the electric charge density induced by Goldstone fields 
\begin{equation}
\delta\rho(\vec k)=\sum_a\kappa_a(k)\phi_a(\vec k),
\end{equation}
we have
\begin{equation}
\kappa_a(k)=-\frac{1}{2}eT\sum_n\int\frac{d^2\vec p}{(2\pi)^3}
{\rm Tr}\gamma_0Q{\cal S}_M(\omega_n,\vec p+\vec k)\gamma_5
G_a{\cal S}_M(\omega_n,\vec p),
\label{kappaa}
\end{equation}
where $Q=\rho_3(a+b\tau_3)$, the matrix $G_4=\varepsilon^1\rho_1$, $G_5=\varepsilon^1\rho_2$, 
$G_6=\varepsilon^2\rho_1$, $G_7=\varepsilon^2\rho_2$
and $G_8=\varepsilon^3\rho_1$ for the five Goldstone fields.

With respect to r, g, and b color indexes, the propagator assumes the block structure
\begin{equation}
{\cal S}_M=\left(\begin{array}{cc}
{\cal S}_M^{SU(2)} & 0 \\
0 & {\cal S}_M^{U(1)} \end{array}\right)
\end{equation}
while the matrix $\varepsilon^\rho$ ($\rho=1,2)$ takes the form
\begin{equation}
\varepsilon^\rho=\left(\begin{array}{cc} 0 & \chi_\rho(2\times 1)\\
-\chi^\dagger(1\times 2) & 0 \end{array}\right).
\end{equation}
Therefore the trace with respect to the group indexes vanishes for $\varepsilon^1$ and 
$\varepsilon^2$. Consequently
\begin{equation}
\kappa_4(k)=\kappa_5(k)=\kappa_6(k)=\kappa_7(k)=0.
\end{equation}

Coming to $\kappa_8(k)$, we firstly take the transpose of the matrix under the trace on RHS 
of (\ref{kappaa}),
then we employ the identity
\begin{equation}
\gamma_0C{\cal S}_M^T(p_0,\vec p)C^\dagger\gamma_0
=-{\cal S}_M(p_0,-\vec p),
\end{equation}
and transform the integration variable from $\vec p$ to $-\vec p-\vec k$. We find
\begin{equation}
\kappa_8(k)=-\kappa_8(k)=0.
\end{equation}

\end{document}